\documentclass[iop]{emulateapj}
\usepackage{graphicx}

\newcommand{\thisstar}{HD\,185351}

\newcommand{\kep}{{\it Kepler}}
\newcommand{\hipp}{{\it Hipparcos}}
\newcommand{\logg}{$\log{g}$}
\newcommand{\teff}{$T_{\rm eff}$}
\newcommand{\msun}{$M_\odot$}
\newcommand{\rhosun}{$\rho_\odot$}
\newcommand{\rhostar}{$\rho_\star$}
\newcommand{\mstar}{$M_\star$}
\newcommand{\rstar}{$R_\star$}
\newcommand{\fbol}{$F_{\rm bol}$}
\newcommand{\rsun}{$R_\odot$}
\newcommand{\ms}{m~s$^{-1}$}

\newcommand{\sfe}{+0.16}
\newcommand{\usfe}{0.04}
\newcommand{\slogg}{3.31}
\newcommand{\uslogg}{0.06}
\newcommand{\steff}{5016}
\newcommand{\usteff}{44}
\newcommand{\smstar}{1.87}
\newcommand{\usmstar}{0.07}
\newcommand{\srstar}{5.07}
\newcommand{\usrstar}{0.16}
\newcommand{\srhostar}{0.014}
\newcommand{\usrhostar}{0.004}

\newcommand{\nmax}{229.8}
\newcommand{\snmax}{6.0}
\newcommand{\delnu}{15.4}
\newcommand{\sdelnu}{0.2}

\newcommand{\iteff}{5042}
\newcommand{\uiteff}{32}
\newcommand{\irstar}{4.97}
\newcommand{\uirstar}{0.07}
\newcommand{\angdiam}{1.132}
\newcommand{\uangdiam}{0.012}

\newcommand{\amstar}{1.99}
\newcommand{\uamstar}{0.23}
\newcommand{\arstar}{5.35}
\newcommand{\uarstar}{0.20}
\newcommand{\arhostar}{0.0130}
\newcommand{\uarhostar}{0.0003}
\newcommand{\alogg}{3.280}
\newcommand{\ualogg}{0.011}

\newcommand{\aimstar}{1.60}
\newcommand{\uaimstar}{0.08}

\newcommand{\asmstar}{1.90}
\newcommand{\uasmstar}{0.15}
\newcommand{\asrstar}{5.27}
\newcommand{\uasrstar}{0.15}
\newcommand{\asrhostar}{0.0130}
\newcommand{\uasrhostar}{0.0003}
\newcommand{\aslogg}{3.273}
\newcommand{\uaslogg}{0.014}

\newcommand{\numax}{\mbox{$\nu_{\rm max}$}}
\newcommand{\Dnu}{\mbox{$\Delta \nu$}}

\shorttitle{Mass and Radius of \thisstar}
\shortauthors{Johnson et al.}

\begin{document}

\title{The Physical Parameters of the Retired A Star \thisstar
}

\author{
John Asher Johnson\altaffilmark{1},
Daniel Huber\altaffilmark{2,3},
Tabetha Boyajian\altaffilmark{4},
John M. Brewer\altaffilmark{4},
Timothy R. White\altaffilmark{5,6},\\
Kaspar von Braun\altaffilmark{7},
Vicente Maestro\altaffilmark{6},
Dennis Stello\altaffilmark{6,8} and
Thomas Barclay\altaffilmark{2,9}
}

\altaffiltext{1}{Harvard-Smithsonian Center for Astrophysics, 60 Garden Street, Cambridge, Massachusetts 02138 USA; Institute for Theory and Computation;}
\altaffiltext{2}{NASA Ames Research Center, Moffett Field, CA 94035, USA}
\altaffiltext{3}{SETI Institute, 189 Bernardo Avenue, Mountain View, CA 94043, USA}
\altaffiltext{4}{Department of Astronomy, Yale University, New Haven, CT 06511, USA}
\altaffiltext{5}{Institut f\"{u}r Astrophysik, Georg-August-Universit\"{a}t G\"{o}ttingen, Friedrich-Hund-Platz 1, 37077 G\"{o}ttingen, Germany}
\altaffiltext{6}{Sydney Institute for Astronomy, School of Physics, University of Sydney, NSW 2006, Australia}
\altaffiltext{7}{Max Planck Institute for Astronomy, K\"{o}nigstuhl 17, 69117 Heidelberg, Germany}
\altaffiltext{8}{Stellar Astrophysics Centre, Department of Physics and Astronomy, Aarhus University, Ny Munkegade 120, DK-8000 Aarhus C, Denmark}
\altaffiltext{9}{Bay Area Environmental Research Institute, 596 1st Street West, Sonoma, CA 95476, USA}

\begin{abstract}
We report here an analysis of the physical stellar parameters of the giant star \thisstar\ using \kep\ short-cadence photometry, optical and near infrared interferometry from CHARA, and high-resolution spectroscopy. Asteroseismic oscillations detected in the \kep\ short-cadence photometry 
combined with an effective temperature calculated from the interferometric angular diameter and bolometric flux
yield a mean density, $\rho_\star = \arhostar \pm \uarhostar$~$\rho_\odot$ and surface gravity, $\log{g} = \alogg \pm \ualogg$. Combining the gravity and density we find $R_\star = \arstar \pm \uarstar$~\rsun\ and $M_\star = \amstar \pm \uamstar$~\msun. The trigonometric parallax and CHARA angular diameter give a radius $R_\star = \irstar \pm \uirstar$~\rsun. This smaller radius, when combined with the mean stellar density, corresponds to a stellar mass $\aimstar \pm \uaimstar$~\msun, which is smaller than the asteroseismic mass by 1.6--$\sigma$. We find that a larger mass is supported by the
observation of mixed modes in our high-precision photometry, the spacing of which is consistent only for $M_\star \gtrsim 1.8$~\msun. Our various and independent mass measurements can be compared to the mass measured from interpolating the spectroscopic parameters onto stellar evolution models, which yields a model-based mass $M_{\star,\rm model} = \smstar \pm \usmstar$~\msun. This mass agrees well with the asteroseismic value, but is 2.6--$\sigma$ higher than the mass from the combination of asteroseismology and interferometry. The discrepancy motivates future studies with a larger sample of giant stars. However, all of our mass measurements are consistent with \thisstar\  having a mass in excess of 1.5~\msun.
\end{abstract}



\section{Introduction}

\thisstar\ ($=$\,KIC\,8566020, HR\,7468, HIP\,96459) is the third brightest target star in the field of view of the NASA \kep\ Mission \citep{kep, basri05, kic}. With a \kep-band magnitude $K_P = 5.034$ ($V = 5.18$), only CH~Cyg and $\theta$~Cyg are brighter. Having exhausted its core hydrogen fuel source \thisstar\ has evolved away from the main sequence and now resides at the base of the red giant branch of the H--R diagram. The \hipp\ catalog lists $B-V = 0.928$, absolute V-band magnitude $M_V = 2.13$ and a parallax-based distance of $40.83 \pm 0.36$~pc \citep{hipp,hipp2}. The \citet{keenan89} catalog of revised MK spectral types classifies \thisstar\ as a G8.5\,III, indicating a giant luminosity class. However, it's location in the observational H--R diagram is consistent with being a class IV subgiant according to the conventions used by \citet{sandage03}, and it is among the ``subgiant'' targets of the Doppler-based planet survey of \citet{johnson06b} and \citet{johnson11}. 

Recent spectroscopic analyses give mass estimates ranging from 1.4--1.7~\msun\ \citep{allende, wang11}, indicating that \thisstar\ was once an F-- or A--type dwarf similar to Procyon or Sirius while on the main sequence---a massive, evolved class of stars that \citet{johnson07a} termed the ``retired A stars.'' However, these and other mass estimates for single stars are based on stellar evolution models, which may contain systematic errors due to, e.g., uncertainties in the treatment of convection and errors related to the assumption of local thermodynamic equilibrium (LTE) in modeling their stellar spectra. Indeed, the mass estimates of subgiants in particular have recently been called into question based on theoretical grounds \citep{lloyd11, lloyd13}, and on the basis of comparing the galactic space motions of evolved and unevolved stars of various masses \citep{schlaufman13}. These studies suggest that, in a statistical sense, subgiants with masses in excess of 1.5~\msun\ should be rare in the Solar Neighborhood. Following this argument, stars like \thisstar\ are much more likely to be the evolved counterparts of G-- or F--type stars, with masses in the range 1.1--1.3~\msun, rather than the elder brethren of A--type stars.

The resolution of this question has important implications for the reliability of stellar evolution models along the subgiant and giant branches. The issue also impacts our understanding of planet occurrence as a function of stellar mass because much of what is known about planets around stars with \mstar~$\gtrsim 1.3$~\msun\ comes from Doppler surveys of evolved stars \citep[e.g.][]{frink02,sato03,hatzes03,johnson07a,nied07}. This is because main-sequence A-- and F--type dwarfs are rapid rotators and exhibit large amounts of radial velocity ``jitter,'' making the detection of even Jovian-mass planets difficult or impossible \citep{galland05}. However, once these stars evolve off of the main sequence, they experience rapid spin-down due to the onset of surface convective layers, which generate magnetic dynamos that carry angular momentum via stellar winds to the Alfv\`en point \citep[e.g.][]{gray85,donascimento00}. 

Surveys of massive, evolved stars have discovered giant planets orbiting evolved stars with masses in excess of $\approx1.4$~\msun\ at rates that are much higher than have been found for solar-mass and M-type dwarf stars, revealing an apparent correlation between stellar mass and giant planet occurrence \citep{johnson07b,lovis07,johnson10c,bowler10}. This relationship has provided important clues about the planet formation process \citep{laughlin04, kennedy08} and hinted at fertile hunting grounds for additional planets via, e.g., high-contrast, direct imaging of main-sequence A-type stars \citep{marois08, lagrange10, crepp11, nielsen13}. However, the reality of the apparent correlation between stellar mass and giant planet occurrence hinges on accurate knowledge of the masses of evolved stars \citep{lloyd13, johnson13}.

In the present work we address this question using several independent and complementary methods to measure the mass of the putative retired A star, \thisstar. Our methodology is similar to the study of the physical properties of the planet-hosting giant stars $\iota$~Draconis and $\beta$~Geminorum \citep{zech08, baines11, hatzes12}. We take advantage of the proximity of HD\,185351 to the Sun and its relatively large physical size ($R_\star \approx 5$~\rsun) to measure its angular diameter using optical and near infrared (NIR) interferometry \citep[see e.g.][]{boyajian13a}. We also leverage the star's placement in the \kep\ field to measure its surface gravity and mean density based on its p-mode oscillation spectrum using \kep\ short-cadence photometry. Using asteroseismic scaling relations extrapolated from the Sun, as has been done for evolved stars by, e.g., \citet{huber13}, we obtain accurate and precise measurements of the stellar mass and radius. We then fit spectra to the star's broad-band spectral energy distribution, along with the interferometric angular diameter, to derive the star's effective temperature. Finally, we compare these independently-measured physical properties to the quantities estimated from the interpolation of the star's spectroscopic properties onto stellar evolution model grids.

\section{Observations and Analysis}

\subsection{Spectroscopy}
\label{sec:spectroscopy}

The relative radial velocity of \thisstar\ has been monitored at high precision ($\sigma_{\rm RV} \approx 5$~\ms) over the past decade as part of the Doppler survey of subgiants performed by \citet{johnson06b}, initiated at the Lick Observatory in Northern California using the Hamilton Spectrometer. The radial velocities of the target stars in this survey are measured with respect to co-added, iodine-free ``template'' spectra \citep{johnson06a}. These template spectra are also useful for measuring the spectroscopic properties of the targets stars. Our Lick template spectra were observed with a resolving power $R = \lambda / \Delta\lambda \approx 50,000$, and a signal-to-noise ratio, S/N~$\approx 130$ at 550~nm. In addition to the two Lick/Hamilton template spectra of HD\,185351, we also obtained three additional templates using the HIgh Resolution Echelle Spectrometer (HIRES) on the Keck 10-meter telescope atop Mauna Kea in Hawaii \citep{vogt94}. Our Keck/HIRES template spectra have $R \approx 55,000$ and S/N~$\approx 240$ at 550~nm.

We derived the global spectroscopic parameters from our high-resolution spectra using an iterative version of the Spectroscopy Made Easy (SME) analysis package \citep{valenti96, vf05} which uses the Yonsei-Yale model grids \citep[Y2;][]{y2} to provide a constraint on surface gravity \citep[cf Figure 1 of][for a illustrative flow chart of the iterative scheme]{valenti09}.  This process helps to break the degeneracies between \logg, \teff, and [Fe/H] \citep{torres12}. The line list used in the analysis is from \citet{vf05} and includes the Mg~b triplet region in addition to spectral segments spanning $\approx 150$~\AA\ between 6000~\AA\ and 6200~\AA. 

The spectra are first analyzed with surface gravity, effective temperature, projected rotational velocity, overall metallicity [M/H], and five elemental abundances (Na, Si, Ti, Fe, and Ni) as free parameters.  Solar values are used for initial parameters except for \teff\ and \logg\ where rough estimates are derived from the star's B-V color.  SME uses forward modelling of the selected spectral region and $\chi^2$ minimization to find the best model. The {\em Hipparcos} distance and the bolometric correction for the star are used to determine $L_\star$ and combined with the spectroscopic \teff, [Fe/H], and [Si/Fe] (as a proxy for [$\alpha$/Fe]) to interpolate in the Y$^2$ evolution grid\footnote{In the analysis described later we make use of the BaSTI model grids. We have confirmed that using the BaSTI grids in our iterative SME analysis yields the same result as using the Y$^2$ grids.}.  If the spectroscopic and model-grid gravities don't agree to within 0.001 dex, the gravity is fixed to that of the grid and the SME analysis is run again.  This process iterates until the surface gravity of the LTE fit converges to the prediction of the evolution models.

Our SME analysis results in a metallicity [Fe/H]~$ = \sfe \pm 0.03$, surface gravity $\log{g} = \slogg \pm 0.06$, and effective temperature \teff~$ = \steff \pm 44$~K. The other spectroscopic parameters are summarized in Table~\ref{tab:concordance}. Note that the errors are the formal uncertainties derived from the SME fitting procedure as described by \citet{vf05}, and as such do not include various unknown systematic contributions. We find that fixing the surface gravity at a value $0.03$~dex lower than our best-fitting value (the difference between our SME-based \logg\ and the value measured from interferometry in \S~\ref{sec:astero}) results in a 9~K lower  effective temperature, and a 0.04~dex lower metallicity. \citet{vf05} identify a possible 0.05~dex systematic error in metallicity from SME, which will affect stellar mass estimates based on interpolating stellar evolution model grids. A systematic error of this size results in a $\pm0.05$~\msun\ change ($\sim2$\%) in the mass measured from model grids. Thus, systematic uncertainties of the same magnitude as our internal errors do not have a large affect on our mass measurements that rely on our SME-based spectroscopic properties. 

\subsection{CHARA Interferometry}
\label{sec:chara}

We obtained long-baseline, optical/near-infrared (NIR) interferometric observations of \thisstar\ using the Center for High Angular Resolution Astronomy (CHARA) Array located at Mount Wilson Observatory, near Los Angeles, CA.  The CHARA Array consists of six 1-meter telescopes in a Y-configuration joined together in a central location on the observatory grounds. The longest available baseline is 330~m, making it the largest effective aperture in the world at optical and NIR wavelengths.

\begin{figure}
\begin{center}
\resizebox{\hsize}{!}{\includegraphics{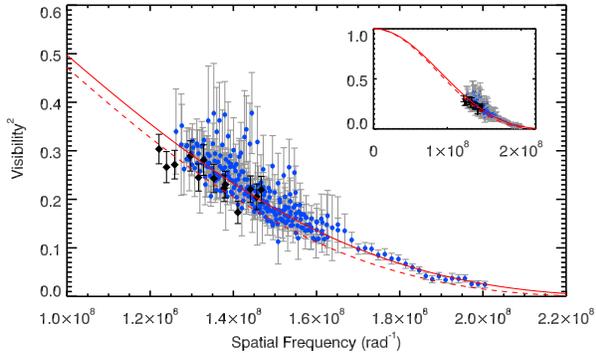}}
\caption{Visibility versus spatial frequency. Black diamonds are measurements made with CHARA Classic, blue circles are from PAVO. The red lines show the fitted limb-darkened model to the combined data. The dashed line is for $\mu = 0.32 \pm 0.04$ (Classic), and the solid line is for $\mu = 0.64 \pm 0.03$ (PAVO). The inset shows the visibility curve over a wider scale. }
\label{fig:visibilities}
\end{center}
\end{figure}

We use a combination of both the CHARA Classic \citep{tenbrummelaar05} and Precision Astronomical Visible Observations (PAVO) beam combiners \citep{ireland08}.  The CHARA Classic instrument is a pupil-plane beam combiner operating in NIR $J$, $H$, and $K^{\prime}$ bands in either two-- or three--telescope configurations \citep{sturmann10}, and can observe objects as faint as $K^{\prime} \approx 9.5$. The PAVO instrument is also a three-beam pupil-plane beam combiner, operating over a wavelength range of 0.65--0.80\,$\mu$m \citep[][ approximately the Bessell $R$-band]{ireland08}, and has a limiting magnitude of $R \lesssim 8$.

The available angular resolution of an interferometer is dependent on the baseline as well as the wavelength.  Thus, at a fixed baseline the PAVO instrument has higher angular resolution compared to Classic owing to the shorter observed wavelengths. However, in practice it is typical to configure the Array depending on the instrument and choice baseline configuration in order to sample the spatial frequencies and UV space appropriate for the science target.   If the goal is to measure a diameter of a symmetric object such as \thisstar, we ensure proper sampling of the visibility curve given observations with longer baselines in the NIR with Classic and shorter baselines in the visible with PAVO.  If resolution is not a necessity, it is advantageous for observations to be made in the infrared where limb-darkening corrections and their associated uncertainties are relatively small compared to those at optical wavelengths.

Both PAVO and Classic are routinely used to measure sub-milliarcsecond (mas) angular diameters of stars \citep{bazot11, derekas11, huber12, maestro13, white13, boyajian13a, vonbraun11a, vonbraun11b, vonbraun12, vonbraun14}. Calibrated visibilities have shown excellent agreement with measurements from various interferometers using independent beam combiners operating in the visible or near infrared \citep{white13, boyajian12b}.

Our CHARA Classic observations were obtained in August of 2012 using the S2-W1 pair of telescopes, which have a maximum baseline of $B_{\rm max} = 249.4$~m in $H$-band \citep[$\lambda = 1.67 \mu$m;][]{tenbrummelaar05}. Our PAVO observations were obtained on August 11, 2012, July 7, 2013,  and April 6--7, 2014 with the W1-W2 pair of telescopes ($B_{\rm max} = 107.9$~m), and on April 10 2014 with the E2-W2 pair of telescopes ($B_{\rm max} = 156.3$~m), in 23 independent wavelength channels between $0.65-0.8$\,nm.  A log of the observations can be found in Table~\ref{tab:interflog}, where we list the UT date, interferometer configuration, number of bracketed observations, and the calibrator stars observed.

We follow the same observing procedures outlined in \citet{boyajian12a, boyajian12b, boyajian13a}, which we briefly summarize herein.  The science star was observed in bracketed sequences along with calibrators stars.  Calibrators were chosen to be unresolved sources that lie within a few degrees on the sky to the science target. In order to select suitable calibrators, we use the {\tt SearchCal} tool developed by the JMMC Working Group \citep{bonneau06, bonneau11}. We investigate each calibrator for any unexpected variance by comparing the data with the other calibrators observed on each night and found none.  Observations were  collected over the course of several nights, rotating between selected calibrators in order to reduce any night-to-night systematics, though we did not identify any evidence of systematic errors within the data set. The August 2012 PAVO observations were taken with only one calibrator, HD\,188665, which has previously been tested and used as a good calibration source with PAVO observations of \kep\ stars in the field \citep{white13}.

To measure the angular diameter of \thisstar\ we fitted a limb-darkened disk model to the calibrated visibility measurements\footnote{Specifically, we  measure the diameter of the Rosseland, or mean, radiating surface of the star. While our result depends on a model-dependent prescription of the limb--darkening, uncertainties in limb-darkening coefficients contribute to the total error budget are an order of magnitude smaller than other error contributions in our measurements \citep[cf \S~2.1 of][]{vonbraun14}.} \citep{hanburybrown74},

\begin{eqnarray}
V &=& \left( \frac{1-\mu_\lambda}{2} + \frac{\mu_\lambda}{3} \right)^{-1} \nonumber \\ 
&\times& \left[ (1-\mu_\lambda) \frac{J_1(x)}{x} + \mu_\lambda \left(\frac{\pi}{2}\right)^{1/2} \frac{J_{3/2}(x)}{x^{3/2}} \right],\label{eqn:vis}
\end{eqnarray}

\noindent where $V$ is the visibility, and $\mu_\lambda$ is the linear limb-darkening coefficient. $J_n(x)$ is the $n^\mathrm{th}$
order Bessel function, and is a function of $x = \pi \theta_\mathrm{LD} B \lambda^{-1}$, where $B$ is the projected baseline, $\theta_\mathrm{LD}$ is the angular diameter after correction for
limb-darkening, and $\lambda$ is the wavelength at which the observations was made. The quantity $B\lambda^{-1}$ is also known as the spatial frequency. The linear limb-darkening coefficients were determined in $H$ and $R$ bands by interpolating the model grid by \citet{claret11} to the spectroscopic measurements of [Fe/H], \logg\ and \teff\ given in \S~\ref{sec:spectroscopy}.
We note that the uncertainties in the limb-darkening coefficients are small compared to the total uncertainty in the angular diameter \citep{huber12,vonbraun14}. Furthermore, oblateness due to rotation is expected to have negligible influence for a slowly rotating evolved star such as HD\,185351.

The model--fitting procedure and parameter uncertainty estimation was performed using the method outlined in \citet{derekas11}, which involves Monte Carlo simulations taking into account uncertainties in the data, wavelength calibration, calibrator sizes and limb-darkening coefficients. A simultaneous fit was made to the Classic and PAVO observations, with a common angular diameter and different limb-darkening coefficients. Figure~\ref{fig:visibilities} shows the observed visibilities and fitted model. We find \thisstar\ has a limb-darkened angular diameter of $\theta_\mathrm{LD} = \angdiam \pm \uangdiam$~mas. Combined with the \hipp\ parallax, this measurement implies a linear radius of \rstar\ $=\irstar \pm \uirstar$~\rsun (corresponding to the Rosseland, or mean, radiating surface of the star). Fitting the Classic and PAVO observations individually provides consistent results (see Table~\ref{tab:interf}).

\subsection{SED Fitting}
\label{sec:sed}

In order to determine \thisstar's effective temperature and luminosity as directly as possible we perform a stellar spectral energy distribution (SED) fit to literature broad-band and spectro-photometric data published in \citet{1963MNRAS.125..557A, 1966MNRAS.133..475A, 1966ArA.....4..137H, 1986EgUBV........0M, 1975MNRAS.172..667J, 1976A&AS...26..275R, 1988iras....1.....B, 2003tmc..book.....C, 1981PDAO...15..439M, 1987A&AS...68..259H, 1972VA.....14...13G, 1991TrSht..63....4K, 1968tcpn.book.....E, 1985BCrAO..66..152B, 2004ApJS..154..673S, glushneva}. Our procedure is analogous to that of \citet{vonbraun14}: we perform a $\chi^2$-minimization of a linearly-interpolated SED template based on the G5\,III and G8\,III templates from the \citet{1998PASP..110..863P} library to the aforementioned literature photometry of \thisstar.

\begin{figure}
  \centering
    \includegraphics[scale=0.29]{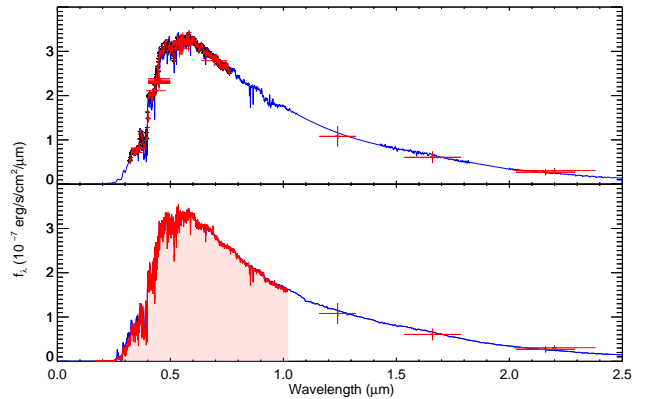}
    \caption{
Top panel: SED fit of spectral templates from the Pickles (1998) library (blue) to photometric measurements of \thisstar\ from the literature (red). Horizontal bars represent the bandwidths of the photometric filters, and the vertical bars represents the literature-based uncertainties, scaled by the corresponding flux values. The 2MASS photometry is saturated for this star \citep{2MASS}, which is evident by the large photometric errors for the points beyond 1~$\mu$m. Bottom panel: SED of \thisstar\ showing a HST STIS spectrum (red, $0.17-1.01\,\mu$m) and 2MASS photometry. The blue line shows the best-fitting ATLAS9 model with solar composition ([Fe/H]=0.0) and microturbulent velocity $\xi=2\,$km\,s$^{-1}$ \citep{castelli04}.}
\label{fig:sed}
\end{figure}

If the literature photometry values are in magnitudes, they are converted to absolute fluxes by application of published or calculated zero points. During the calculation of $\chi^2$ only the central broad-band filter wavelengths are correlated with the SED template's flux value averaged over the filter transmission range in wavelength. Literature spectrophotometry data are used to trace out the shape of the SED in more detail than broadband data, and they thus help in the manual selection of the input spectral template. The SED template is scaled to minimize $\chi^2$ and then integrated over wavelength to obtain the bolometric flux, $F_{\rm bol}$.

In our fitting procedure, the value for interstellar reddening, $A_V$, is allowed to float. The best fit is obtained when $A_V$ is 0, which is sensible given \thisstar's small distance. Based on 325 photometric data points, we calculate the bolometric flux to be $F_{\rm{bol,Pickles}} = 2.751\pm0.013 \times 10^{-7}$~erg~cm$^{-2}$~s$^{-1}$ with a $\chi^2_{red} = 2.75$. To account for uncertainties due to the absolute flux calibration of the photometry, we added a 3\% error in quadrature to the formal uncertainty, yielding $F_{\rm{bol,Pickles}} = 2.751\pm0.084 \times 10^{-7}$~erg~cm$^{-2}$~s$^{-1}$.

We also performed an alternative estimation of $F_{\rm{bol}}$ using a spectrum taken with the Space Telescope Imaging Spectrograph (STIS) aboard the the Hubble Space Telescope (HST)\footnote{\texttt{http://archive.stsci.edu/prepds/stisngsl/}}. The STIS spectrum covers the $0.17-1.01\,\mu$m wavelength range at an intermediate spectral resolution of $R\sim1000$, corresponding to $\simeq60\,\%$ of the total radiated power of \thisstar. Potential errors induced by absolute flux calibration of ground-based spectro-photometry are reduced considerably by using this approach, as the uncertainties of the absolute flux calibration of the STIS spectrum are $\leq 1\,\%$ \citep{Bohlin04}.

The bolometric flux is measured by fitting, through $\chi^2$ minimization, the STIS data to theoretical atmosphere models interpolated from a grid of ATLAS9 models with solar composition ([Fe/H]=0.0) and microturbulent velocity $\xi=2\,$km\,s$^{-1}$. The parameters of the model are the atmosphere temperature $T$ and \fbol. Similar to the Pickles (1998) fit the best fit is found for $A_V=0$, as expected for such a nearby star. The bottom panel of Figure~\ref{fig:sed} shows the best-fitting model. We have confirmed the validity of the fit longward of $\lambda=1.01\,\mu$m by comparing the 2-MASS near-infrared fluxes from $J$, $H$ and $K_s$ broadband photometry with the fitted values. The resulting \fbol\ is $F_{\rm{bol,STIS}} = 2.76\pm0.04 \times 10^{-7} $~erg~cm$^{-2}$~s$^{-1}$, computed as the sum of the total flux of the STIS spectrum ($1.6376\pm0.0035\times 10^{-7} $~erg~cm$^{-2}$~s$^{-1}$) and the best-fit SED integrated over wavelength outside the $0.17-1.01\,\mu$m range ($1.120\pm0.034\times10^{-7} $~erg~cm$^{-2}$~s$^{-1})$.

Both estimates of the bolometric flux for \thisstar\ are in excellent agreement. We calculate our final estimate of the bolometric flux for \thisstar\ as the weighted average of $F_{\rm{bol,Pickles}}$ and $F_{\rm{bol,STIS}}$, yielding $F_{\rm{bol}}=2.758\pm0.036 \times 10^{-7} $~erg~cm$^{-2}$~s$^{-1}$.

\subsection{{\emph Kepler} Asteroseismology}
\label{sec:astero}

\subsubsection{Background}
\label{sec:background}
Photometric measurements of the integrated flux from a star of sufficient precision reveal brightness oscillations from p-modes driven by stochastic convective motion near the stellar photosphere. This convective motion drives standing waves within the star characterized by spherical degree $l$ (the total number of surface nodes), azimuthal order $m$ (the number of nodes along the stellar equator), and radial order $n$ (the number of nodes from the center to the surface of the star). Modes with low $l$ and high $n$ can be observed in the stellar flux integrated over the visible stellar surface, and the nature of these modes is related to the star's fundamental physical characteristics.

The competition between convective driving and damping in the star's surface layers gives rise to an envelope of frequencies in the photometric power spectrum characterized by the frequency of maximum power, \numax, and the large frequency separation, \Dnu. The latter is the average separation between power-spectrum peaks with the same value of $l$ and consecutive values of $n$. The first-order asymptotic analysis of p-mode oscillations shows that $\Delta \nu \propto \rho_\star^{1/2}$, where \rhostar\ is the mean stellar density \citep{ulrich86}. Scaling with respect to the Solar p-mode spectrum gives

\begin{equation}
\Delta \nu = \Delta\nu_{\sun} \left(\frac{M_\star}{M_\odot}\right)^{1/2} \left(\frac{R_\star}{R_\odot}\right)^{-3/2}.
\label{eqn:dnu}
\end{equation}

The frequency of maximum power, $\nu_{\rm max}$, has been proposed to scale with the acoustic cut-off frequency, $\nu_{\rm ac}$, which is proportional to the inverse of the dynamical timescale or $\nu_{\rm max} \propto \nu_{\rm ac} \propto c_s/H_p$ \citep{brown91,kjeldsen95}. Here, $c_s$ is the adiabatic sound speed in the star's photosphere, and $H_p$ is the photosphere's pressure scale height. For an ideal gas, $c_s \propto T^{1/2}$, where $T$ is the mean stellar temperature in the photosphere, and the scale height is given by $H_p \propto T/g$, where $g$ is the surface gravity. Making use of homology relations and scaling with respect to the Sun yields \citep{kjeldsen95,belkacem11}:

\begin{equation}
\nu_{\rm max}  = \nu_{\rm max,\sun}~\left(\frac{M_\star}{M_\odot}\right) \left(\frac{R_\star}{R_\odot}\right)^{-2} \left(\frac{T_{\rm eff}}{5777\,K}\right)^{-1/2}.
\label{eqn:numax}
\end{equation}

Equations \ref{eqn:dnu} and \ref{eqn:numax} are approximate relations and require 
careful calibration, in particular for low-luminosity RGB stars such as 
\thisstar, which are significantly more evolved than 
the Sun. Detailed reviews of theoretical and empirical tests of asteroseismic 
scaling relations can be found in \citet{belkacem12} and \citet{miglio13}, and we 
present a brief discussion as relevant for \thisstar\ here.

Empirical tests using long-baseline interferometry and Hipparcos parallaxes have shown 
that asteroseismic radii 
calculated from scaling relations are accurate to $\lesssim$4\% for main-sequence and 
subgiant stars \citep{huber12, silva12}. While such comparisons are 
currently still limited to $\sim15\%$ for RGB stars due to poor parallax precisions, 
empirical tests using cluster giants (including low-luminosity RGB stars 
similar to HD185351) have shown agreement within $5\%$ in radius \citep{miglio12}. 
Empirical tests of asteroseismic masses are more challenging, and
have relied on eclipsing binaries and cluster members. 
\citet{miglio12} showed that masses for RGB stars in 
NGC6819 show no systematic offset with a scatter of $\lesssim$\,15\%, while the average 
asteroseismic cluster mass lies within $\sim$\,7\% of the mass determined 
from near turn-off eclipsing binary stars \citep{brogaard12}. 

More recently, 
\citet{frandsen13} measured the mass 
of an oscillating red giant in a double-lined eclipsing binary, which 
was found to be $\sim10-15$\% more massive than the seismic mass \citep{hekker11,huber14}. However, 
it is likely that the giant is a He-core burning red clump star, for which systematic 
offsets in the $\Dnu$ scaling relation have been noted  \citep{miglio12}. 
Additional, yet model-dependent tests can be performed by comparing 
properties derived from detailed modeling of individual oscillation 
frequencies, which contain information on the core properties of the star, such as the sound 
speed gradient.
Such detailed modeling efforts for RGB stars have yielded radii and masses that agree
within $\sim$3\% and $\sim$5\% of the values derived from scaling relations 
\citep{mosser10,dimauro11,jiang11,huber13b}.

In summary, various tests to date have shown that 
asteroseismic scaling relations for stars in similar evolutionary stages to \thisstar\
can be expected to be accurate to $\sim5$\% and $\sim$10\% in radius and mass, 
respectively. We note that improvements to the \Dnu\ scaling relation based on 
models have been proposed in the 
literature, for example based on the comparison of \Dnu\ calculated from individual 
frequencies with model densities \citep{white11}, the extension of the asymptotic 
relation to second order \citep{mosser13} or theoretical relations between fundamental 
properties for red giants \citep{wu14}. The effect of these corrections on the derived 
fundamental properties for \thisstar\ are discussed in \S~\ref{sec:scalingpars}.

\subsubsection{Data Preparation}

\begin{figure}
\begin{center}
\resizebox{\hsize}{!}{\includegraphics{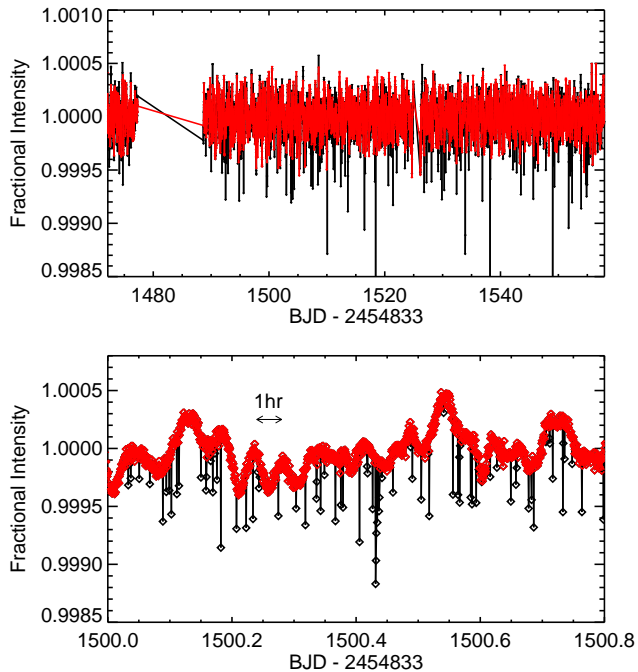}}
\caption{Top panel: Full Quarter 16 short-cadence time series of \thisstar. Black and red data points show the time series before and after the outlier rejection procedure described in the text. Note that only 5\% of all data is shown for clarity. Bottom panel: Same as the top panel but for a 1-day segment. Here, the data are shown with the original 1-minute sampling. Note the $\sim 1$ hour variability in the light curve which is due to oscillations in \thisstar.}
\label{fig:kepdata}
\end{center}
\end{figure}

The \kep\ Mission observed \thisstar\ in long-cadence mode ($\approx 30$~min cadence) from Q1-3, spanning a total of roughly 200~days. However, due to the lack of a proper dedicated pixel mask, and because \thisstar\ is expected to oscillate very close to the long-cadence Nyquist frequency ($\sim 300\mu$Hz), the quality of the photometry is not amenable for asteroseismology. 
In order to detect oscillations in \thisstar\ with sufficient sampling and signal-to-noise ratio (S/N), we obtained one quarter of Kepler short-cadence data with a dedicated pixel mask through an application for Director's Discretionary Time (DDT) in Quarter 16. The data set spans a total of 85.6 days, with a $\sim10$ day gap during the first month due to a spacecraft safe mode (see Fig~\ref{fig:kepdata}).

Inspection of the raw data of \thisstar\ showed a considerable number of outliers below the average flux level. We attribute these outliers to the increased pointing jitter during Q16\footnote{These data were obtained shortly before a reaction wheel failed in Q17. Increased friction that eventually led the reaction wheel failure manifested as stochastic pointing jitter in preceding Quarters.}, causing the photocenter to move sporadically outside the dedicated pixel mask. To reject these outliers we calculated the flux difference of each consecutive data point pair for the full time series. Then, all data points with a flux decrease greater than 3 times the standard deviation of the flux differences over the entire dataset were removed. This procedure was iterated until the residual scatter converged. Finally, we applied a Savitzky-Golay filter with a width of 2 days to the light curve to remove any instrumental and intrinsic low-frequency variability that could influence the oscillation signal, and applied a 4--$\sigma$ clipping using a 1-day moving mean. The detrended light curve with and without the adopted outlier rejection is shown in Figure \ref{fig:kepdata}.

\subsubsection{Fundamental Properties from Scaling Relations}
\label{sec:scalingpars}

\begin{figure*}
\begin{center}
\resizebox{\hsize}{!}{\includegraphics{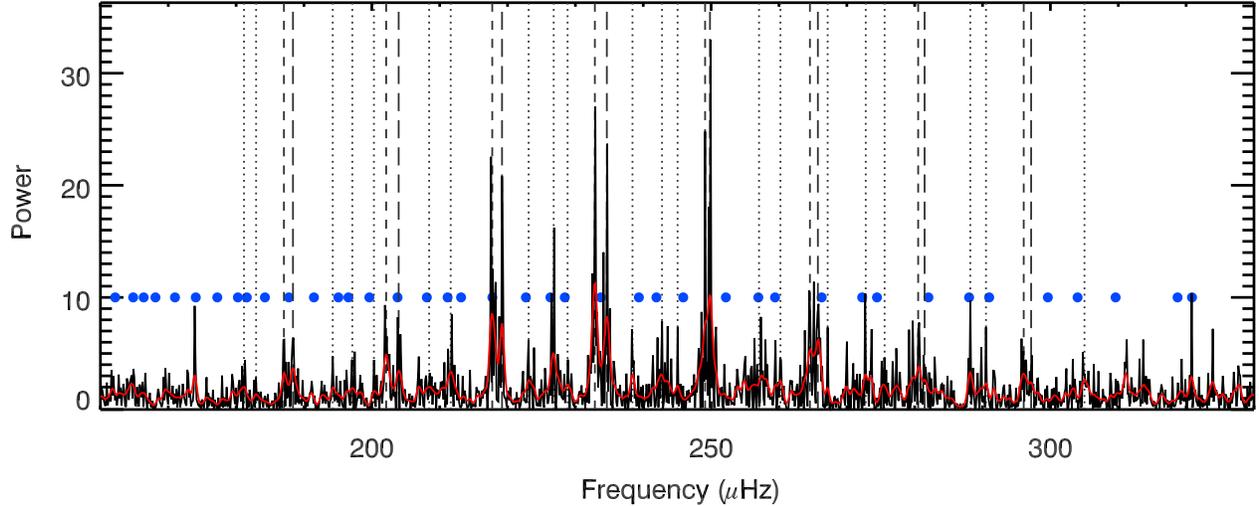}}
\caption{Background-corrected power spectrum of \thisstar, with the smoothed power-spectrum shown in red. Long-dashed, dotted, and short-dashed lines indicate the locations of identified $l$=0, 1, and 2 modes, respectively. Expected locations of l=1 mixed modes from the asymptotic relation are indicated by blue circles.}
\label{fig:powerspectrum}
\end{center}
\end{figure*}

\begin{figure}
\begin{center}
\resizebox{\hsize}{!}{\includegraphics{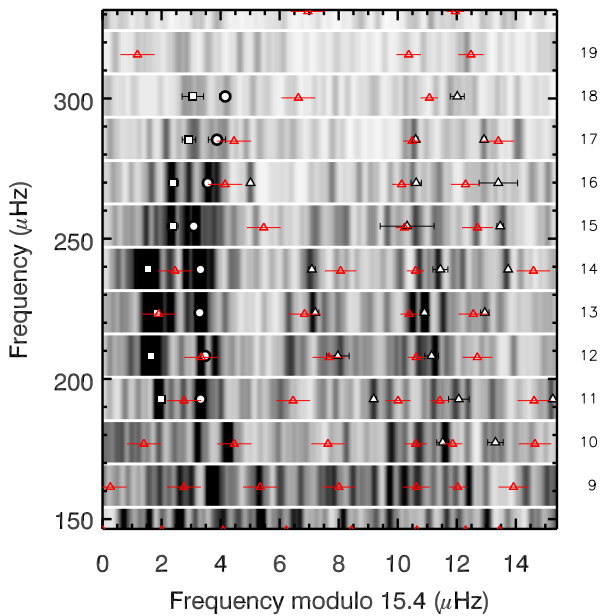}}
\caption{\'Echelle diagram of \thisstar, showing fitted frequencies in white. Modes are identified as $l=0$ (circles), $l=1$ (triangles) and $l=2$ (squares). Expected l=1 mixed modes from the asymptotic relation are shown by the red open triangles, with horizontal bars showing the maximum expected rotational splitting.  For reference, a grey-scale map of the smoothed power spectrum is shown in the background. Numbers to the right of the plot indicate radials order of the $l=0$ modes.}
\label{fig:echelle}
\end{center}
\end{figure}

We calculate the power spectrum from the detrended light curve using the method described by \citet{huber09} to model the background power due to stellar granulation and detect the signature of oscillations. Figure~\ref{fig:powerspectrum} shows the region of the background-corrected power spectrum centered on the detected power excess due to the stellar oscillations. The frequency of maximum power as measured from the smoothed, background--corrected power spectrum is $\nu_{\rm max}= \nmax \pm \snmax~\mu$Hz. To measure the large frequency separation we calculated \'echelle diagrams with trial values of $\Delta \nu$ to align the $l=0$ modes, yielding $\Delta \nu = \delnu \pm \sdelnu~\mu$Hz. 
The \'echelle diagram of \thisstar\ is shown in Figure~\ref{fig:echelle}. Uncertainties on the measured values were estimated from Monte-Carlo simulations performed on synthetic power spectra calculated for a $\chi^2$ distribution with two degrees of freedom, as described in \citet{huber11}. For each synthetic power spectrum the $\Delta \nu$ and $\nu_{\rm max}$ measurement was repeated, and the uncertainties were calculated as the standard deviation of the resulting distributions. We note that the $~1\%$ and $~3\%$ uncertainties on $\Delta \nu$ and $\nu_{\rm max}$ are compatible with typical uncertainties reported in the literature for Kepler observations \citep[e.g.,][]{hekker11b}.

Our measured values of \Dnu\ and \numax\ for \thisstar\ are fully consistent with the relationships between both quantities \citep{stello09,hekker09,mosser10}. Combining \numax\ and \Dnu\ with the interferometric effective temperature derived in \S~\ref{sec:sed} yields \rstar~$= \arstar \pm \uarstar$~\rsun, \mstar~$=\amstar \pm \uamstar$~\msun, \logg~$=\alogg \pm \uaslogg$ and \rhostar~$ = \arhostar \pm \uarhostar$~\rhosun. The adopted solar reference values, which were measured using the same method as applied to HD\,185351, are $\nu_{\rm max,\sun}=3090~\mu$Hz, $\Delta\nu_{\sun}=135.1~\mu$Hz \citep{huber11}.

We note that modifications of the scaling relations suggested in the literature do not significantly change these results. Using stellar mass as an example, the corrections yield \mstar $= 1.89$~\msun\ using the relations by \citet{mosser13}, and \mstar $= 1.97~$\msun\ using the relations by \citet{wu14}. The \Dnu\ correction by \citet{white11} for stars with a temperature similar to \thisstar\ is $<0.1\%$, and hence does not significantly change the stellar mass estimate.

In addition to evaluating Equations \ref{eqn:dnu} and \ref{eqn:numax} directly, \Dnu\ and \numax\ can be used as input values to interpolating evolutionary models. This method has the advantage of yielding smaller formal uncertainties since metallicity information can be taken into account, and unphysical solutions based on evolutionary theory are discarded \citep{gai11}. Combining \teff\ and metallicity derived from the SME analysis with \numax, \Dnu\ and BaSTI evolutionary models \citep{basti} we derived an additional set of asteroseismic properties, yielding  \rstar~$=\asrstar \pm \uasrstar$~\rsun, \mstar~$=\asmstar \pm \uasmstar$~\msun, \logg~$=\aslogg \pm \uaslogg$ and \rhostar~$=\asrhostar \pm \uasrhostar$~\rhosun. Our asteroseismic measurements of the stellar properties based on scaling relations are given in Table~\ref{tab:concordance}.

\subsubsection{Mass Constraints from Mixed--Mode Period Spacings}
\label{sec:mixedmodes}

In evolved stars, the dipole ($l=1$) mixed modes are particularly useful for determining stellar parameters and structure \citep[e.g.][]{metcalfe10,bedding11,mosser11,beck12,benomar12,deheuvels12}. These modes occur when acoustic p-mode oscillations in the outer envelope of the star couple to g-mode (gravity) oscillations in the core \citep{osaki75,aizenman77}. The signature of mixed modes is clear in the \'echelle diagram in Figure~\ref{fig:echelle}, with several dipole modes in each radial order. While p modes of the same degree, $l$, and consecutive radial order, $n$, are approximately equally spaced in frequency, g modes are approximately equally spaced in period \citep{tassoul80}. Due to mode bumping, the observed period spacing of mixed modes will be significantly smaller than the true, underlying g--mode period spacing, $\Delta\Pi_1$. However, by measuring the observed period spacing, a lower limit may be placed on $\Delta\Pi_1$, and if a sufficient number of modes are observed, its value may be deduced. 

According to models, evolved stars show a strong mass dependency on $\Delta\Pi_1$ \citep{white11,stello13}.  While this mass dependency is strongest during the subgiant phase \citep{benomar12,benomar13} the effect is still significant along the red giant branch for stars with non-degenerate or partially-degenerate cores, corresponding to $M \gtrsim 1.8~M_\odot$ \citep{stello13}. Hence, if \thisstar\ does have a high mass, then the period spacing should reflect this.

\begin{figure}
\begin{center}
\resizebox{\hsize}{!}{\includegraphics{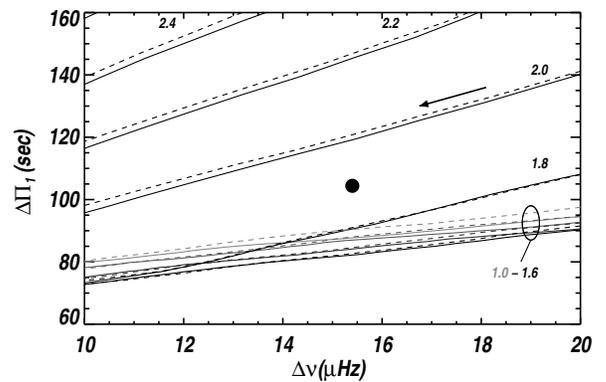}}
\caption{Evolutionary model tracks showing the gravity mode period spacing for dipole modes, $\Delta\Pi_1$, as a function of the large frequency separation, $\Delta\nu$. The mass in solar units is indicated for each track. Solid lines are for models with Fe/H]=+0.0 dex, and dashed lines are for models with [Fe/H]=+0.2 dex. Lower mass tracks ($1.0 < M/M_\odot < 1.6 $) are indicated by a black oval and are shown in greyscale with lower mass indicated by lighter grey. Stars evolve from right to left indicated by the arrow. The location of \thisstar\ is shown by the black circle. The uncertainty in the measurements of $\Delta\Pi_1$ and $\Delta\nu$ is smaller than the symbol size.}
\label{fig:dpidnu}
\end{center}
\end{figure}

To determine the value of $\Delta\Pi_1$ in \thisstar, we first measured the mode frequencies. The background-corrected power spectrum was smoothed by a Gaussian function with full-width at half maximum of $0.4~\mu$Hz, and significant peaks were identified. A global Markov chain Monte Carlo (MCMC) fit was made to the power spectrum with the frequencies, heights and widths of the identified modes as free parameters \citep[e.g.][]{benomar09,handberg11}. The measured mode frequencies and their uncertainties are given in Table~\ref{tab:frequencies}, and indicated in Figures~\ref{fig:powerspectrum}~and~\ref{fig:echelle}. In many red giants observed by \kep, $l=1$ modes are seen to be split by rotation into $m=0, \pm 1$ components \citep{beck12,deheuvels12,mosser12b}. However, no clear indication of this splitting can be found in \thisstar. This may be due to this power spectrum having a relatively low frequency resolution and signal-to-noise ratio compared to other \kep giants. Alternatively, \thisstar\ may have a low inclination, which suppresses the $m=\pm1$ components \citep{gizon03}.

We performed a MCMC fit of the asymptotic relation for mixed modes \citep{mosser12a} to the observed $l=1$ modes to determine p-- and g--mode parameters, assuming that only the $m=0$ component is present. The $l=0$ modes were included in the fit to constrain the p--mode parameters, such as $\Delta\nu$. We find the underlying g--mode period spacing $\Delta\Pi_1 = 104.7 \pm 0.2~\mathrm{s}$. The locations of $l=1$ frequencies predicted from the asymptotic relation are indicated by the blue circles in Figure~\ref{fig:powerspectrum} and red triangles in Figure~\ref{fig:echelle}. 

To investigate the possible impact of undetected rotational splitting on our measured period spacing, we determined the maximum expected rotational splittings following the results of \citet{mosser12b}. Mixed modes that have a stronger $g$-mode character are more sensitive to the rotation rate in the core, while mixed modes that are dominated by a $p$-mode character are more sensitive to the envelope. The cores of red giants rotate substantially faster than their envelopes, and so the observed frequency splitting increases with $g$-mode characteristics. \citet{mosser12b} empirically described this variation in rotational splitting with a Lorentzian profile. For stars of a similar evolutionary state to \thisstar, they found the maximum rotational splitting in a star to vary between 0.2~and~0.6~$\mu$Hz. Taking 0.6~$\mu$Hz as a maximum expected rotational splitting for \thisstar, and the typical values \citet{mosser12b} found for the width and amplitude parameters of the Lorentzian profile, we calculated the rotational splittings for each of the asymptotic frequencies. The size of these splittings is shown by the horizontal bars on the red asymptotic frequencies in Figure~\ref{fig:echelle}. The spacing between the mixed modes is significantly larger than the expected rotational splittings, and so we conclude that the non-detection of rotational splittings has not impacted on our determination of the period spacing.

Figure~\ref{fig:dpidnu} shows $\Delta\Pi_1$ for \thisstar\ relative to a grid of models from
\citet[][solid]{stello13} supplemented by a grid of super solar metallicity ([Fe/H]~$+0.2$; dashed) that bracket the value of \thisstar. The models were generated using the MESA {\tt 1M\_pre\_ms\_to\_wd} test suite \citep{paxton11,paxton13}. MESA derives $\Delta\Pi_1$ from the integral of the buoyancy frequency and derives $\Delta\nu$ from the integral of the sound speed \citep[see][for details]{stello13,paxton13}.
Stars evolve from right to left in this diagram caused by their expansion, and hence decrease in the mean density and $\Delta\nu$. Lower mass tracks ($M_\star < 1.6~M_\odot$) show similar period spacings at a given $\Delta\nu$, while this degeneracy is lifted for more massive stars. We note that this analysis uses the metallicity measurement from SME, but  $\Delta\Pi_1$ is minimally affected by metallicity as can be seen in Figure~\ref{fig:dpidnu}. By matching tracks through the position of \thisstar\ in Figure~\ref{fig:dpidnu} we find it to be  consistent with a mass of 1.85--1.90~\msun. This mass range accounts for the fact that model values of $\Delta\nu$ are based on the integral of the sound speed, which shifts the tracks to the right by up to 3\% relative to the observed value \citep{stello09}.

While the mass estimate based on mixed modes is model-dependent, the period 
spacing probes the conditions in the stellar core and hence provides
valuable independent information compared to other model-dependent mass estimates based 
on atmospheric properties. More detailed modelling of the oscillations, which has been done for other stars \citep[e.g.][]{metcalfe10,dimauro11} but is beyond the scope of this paper, may provide a precise measurement of the age of \thisstar.

\section{Results}
\label{sec:results}

\begin{figure}
\begin{center}
\resizebox{\hsize}{!}{\includegraphics{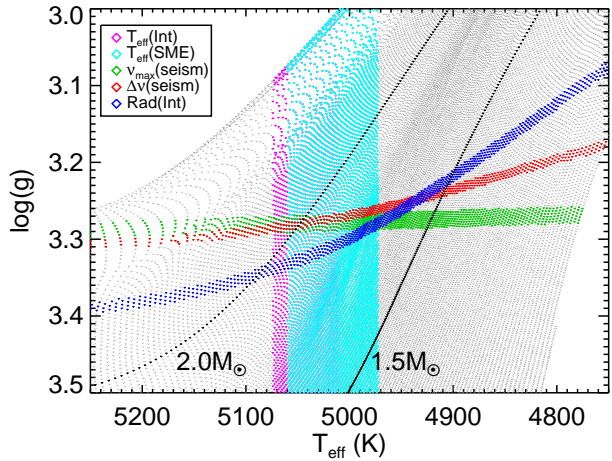}}
\caption{Surface gravity versus temperature for the BaSTI evolutionary tracks \citep{basti} with [Fe/H]$ = +0.16$ \citep{basti}, as measured from our LTE spectral analysis. Colored models show the 1--$\sigma$ constraints from our observations: effective temperature derived from the interferometric angular diameter and bolometric flux (magenta), effective temperature from spectroscopy (cyan), \mstar \rstar$^{-2}$ \teff$^{-0.5}$ from the frequency of maximum power (green), density derived from the large frequency separation (red), and radius from the interferometric angular diameter and \hipp\ parallax (blue). Black lines highlight masses of 1.5~\msun\ and 2.0~\msun, respectively.}
\label{fig:concordance}
\end{center}
\end{figure}

The results of our various independent analyses, and the constraints they place on the mass of \thisstar\ are summarized in Figure~\ref{fig:concordance}. This concordance diagram plots effective temperature versus surface gravity (\logg), at a fixed [Fe/H]$ = \sfe$. The small dots are discrete points sampled from the BaSTI stellar evolution models, which are better sampled and thus better visualized along the subgiant and giant regions of the H--R diagram than are the Y2 models used in \S~\ref{sec:spectroscopy}. The colored bands illustrate the constraints provided by our interferometric, astrometric and spectroscopic analyses, with widths showing the 1--$\sigma$ confidence regions. For example, our \teff\ estimates from SME and our SED fit are shown as vertical magenta and cyan bands, respectively, and have significant overlap, lending confidence that we have derived the temperature of \thisstar\ both accurately and precisely.

Our asteroseismic constraints are shown as red and green, roughly horizontal bands, based on the large separation (\Dnu) and the frequency of maximum oscillation power (\numax). These bands cross at roughly \teff~$=5050$~K, in the region of overlap from the independent measurements of \teff, corresponding to \mstar~$ \approx 2$~\msun.

The final constraint illustrated on this figure is provided by the interferometric measurement of the stellar radius, \rstar~$ = \irstar \pm \uirstar$~\rsun\ (blue). Iso-radius contours in the \logg--\teff\ plane run roughly from lower left to upper right in this diagram. As evident by the position of the blue band with respect to the other constraints, there is some tension at the $2$--$3\sigma$ level between the mass constraint provided by the radius estimate and the asteroseismic and spectroscopic measurements. However, in the region where 
most of our constraints overlap, near \teff~$4980$~K, corresponds to $\sim1.7$~\msun. Thus, all of our independent measurements of the mass of \thisstar\ are {\bf consistent with masses \mstar~$> 1.5$~\msun.}

It should be noted that Figure~\ref{fig:concordance} illustrates our mass constraints with respect to a \emph{theoretical} model grid. One of the primary motivations of our study is to test the accuracy of these types of stellar evolution models using independent measurements. Table~\ref{tab:concordance} lists our estimates of various stellar parameters using different combinations of our measurements. Column~1 lists the full set of stellar parameters, spectroscopic and physical, that we obtain from a combination of our SME spectral analysis and the interpolation of these spectroscopic parameters onto the Y2 stellar evolution model grids. This is the standard technique used to estimate the masses of isolated field stars, as in, e.g., \citet{vf05, takeda07, hekker07, takeda08, johnson07b, johnson13}. These are the parameters that we wish to test with our various methods. In the following subsections we describe the outcome of these comparisons for individual stellar parameters.

\subsection{Effective Temperature}

The effective temperature of \thisstar\ is determined using two approaches. One method, presented in \S~\ref{sec:sed}, makes use of the Stefan-–Boltzmann law, together with our measurements of the parallax-based distance, stellar angular diameter ($\theta$) and bolometric flux (\fbol), which gives \teff~$ = \iteff \pm \uiteff$~K. We consider this method to be empirical in that it 
relies only weakly model-based assumptions (e.g. limb-darkening), and it sidesteps the intricacies and assumptions employed in modeling the observed stellar spectrum, which often relies on the assumption of plane-parallel atmospheres in LTE, among other simplifications. The star's spectral energy distribution is used to estimate the bolometric flux, but we do so using observed stellar spectra of stars with well-measured \fbol, rather than relying on synthetic spectra. However, checking our result based on empirical spectra against model spectra shows close agreement (\S~\ref{sec:sed}).

We find very close agreement between our SED-based measurement of the effective temperature, and that of our SME analysis. This bolsters the reliability of previous estimates of giant and subgiant stellar properties using LTE spectral modeling since systematic errors in \teff\ can lead to large errors in \mstar\ and \rstar\ \citep[e.g.][]{vf05,lloyd13}.

\subsection{Surface Gravity}

Our LTE spectral synthesis modeling is implemented using SME, which models the stellar surface gravity by fitting to the damping wings of the Mg\,II\,b triplet lines \citep{valenti96,vf05}. This method works well for \logg~$\gtrsim 4$, but the uncertainty increases for lower surface gravity due to significant weakening of the Mg\,II\,b damping wings. Also, \logg\ is correlated with both \teff\ and [Fe/H] in SME, which can bias the model-grid-interpolated stellar mass \citep{torres12,huber13}. It is therefore worthwhile to compare our asteroseismic and spectroscopic values of \logg.

As seen in Table~\ref{tab:concordance}, our SME analysis yields \logg~$= \slogg \pm \uslogg$, which compares well with the value obtained from \numax\ (\logg~$= \alogg \pm \ualogg$). Equation~\ref{eqn:numax} shows that the effective temperature enters into the asteroseismic estimate of \logg, but it does so only weakly as \teff$^{-1/2}$, and as a result, errors on \teff\ propagate weakly, as $\frac{1}{2} \sigma_{T_{\rm eff}}^2$. The fractional uncertainty on our SED-based \teff\ is $\sim1$\%, and are therefore negligible compared to the measurement errors in our asteroseismic values of \numax\ and \rstar, which are 2.6\% and 3.8\%, respectively.

Just as with our estimates of \teff, we find close agreement between our empirical measurement and model-based estimate of \logg, which is remarkable given the low surface gravity of \thisstar. This is likely a result of the iterative scheme used in our SME analysis, which differs from the unconstrained SME analysis employed in the critical evaluation of SME in \citet{torres12}.

\subsection{Mean Density}

Our model-grid interpolation of our spectroscopic parameters from SME provide an estimate of the mean stellar density, giving \rhostar~$ = \srhostar \pm \usrhostar$\,g\,cm$^{-3}$. We also estimate \rhostar\ using the large frequency separation observed in our asteroseismic measurements, \Dnu, which gives \rhostar~$= \arhostar \pm \uarhostar$\,g\,cm$^{-1/3}$. We therefore find close agreement between our model-based and empirical estimates of another key physical characteristic of \thisstar.

\subsection{Radius}
\label{sec:radius}
The interpolation of our SME parameters onto stellar evolution model grids also provides an estimate of the stellar radius, giving \rstar~$=\srstar \pm \usrstar$~\rsun. Our interferometric measurements compare well with this value, yielding \rstar~$=\irstar \pm \uirstar$~\rsun, resulting in a 0.6--$\sigma$ agreement (relative to the quadrature-sum of the two errors) between our model-based and empirical measurement.

We also estimate \rstar\ using our asteroseismic measurements of \Dnu\ and \numax, combined with our \teff\ measurement from interferometry, which gives \rstar~$=\arstar \pm \uarstar$~\rsun. This agrees with our model--based estimate to within 1--$\sigma$. Our interferometric and asteroseismic values of \rstar\ bracket our SME$+$model grid value, and the weighted average of our two empirical measurements is \rstar~$=4.98 \pm 0.01$~\rsun, which agrees with the model-based value to  0.56~$\sigma$.

As described in \S~\ref{sec:astero}, we also measured the stellar radius using our asteroseismic measurements under the constraints provided by our SME-based metallicity ([Fe/H]) and the BaSTI stellar evolution models. This gives \rstar~$=\asrstar \pm \uasrstar$~\rsun, which is larger than, but comparable within errors to all of our other measurements.

\subsection{Mass}

The key physical parameter of interest for evolved stars like \thisstar\ is stellar mass. Our model-based estimate, measured using spectral synthesis and model-grid interpolation, is \mstar~$= \smstar \pm \usmstar$~\msun. This value agrees well with our asteroseismic-only measurement made by combining Equations~\ref{eqn:dnu} and \ref{eqn:numax} (\mstar~$= \amstar \pm \uamstar$~\msun). We also find close agreement with the mass measured by interpolating the asteroseismic \logg\ and \rhostar, plus the spectroscopic [Fe/H] and \teff, onto the BaSTI model grids (\mstar~$= \asmstar \pm \uasmstar$~\msun). This second model-based procedure is illustrated in Figure~\ref{fig:concordance} in the region of overlap among the asteroseismic parameters and SED \teff. 

We also estimated the stellar mass by combining the asteroseismic density calculated from the large frequency spacing \Dnu\ (cf Equation~\ref{eqn:dnu}) with the interferometric radius. This is our least model-dependent estimate of the stellar mass in that it is independent of any aspect of our SME spectral analysis and evolution models, which are under scrutiny in this study. We find \mstar~$ = \aimstar \pm \uaimstar$\msun, which is smaller than our SME model-based value by 2.6\,$\sigma$. The value is also smaller than the mass derived from asteroseismic scaling relations (by 1.6\,$\sigma$), asteroseismic scaling relations combined with BaSTI models (by 1.8\,$\sigma$), and the mass implied from the gravity mode period spacing. 

As illustrated in our concordance diagram (Figure 7), the mass difference could be reconciled either by a systematic increase in the interferometric radius (upward shift of blue band) or a systematic increase in \Dnu\ and \numax\ (downward shift of red and green bands). The required offsets in measured quantities are a $\sim$5\% decrease in the parallax, a $\sim$6\% increase in angular diameter, or a 9\% increase in \Dnu\ and \numax. \thisstar\ is a photometric standard star with no indication of a binary companion, and hence a systematic error in the Hipparcos parallax is unlikely. While angular diameters can be affected by systematic calibration errors, the agreement of our estimates using two different instruments and different calibrators rule out a shift as a large as 6\%. Corrections to asteroseismic scaling relations for stars that are more evolved than the Sun have been proposed, but so far theoretical investigations and empirical tests have ruled out offsets as large as 9\% for \Dnu\ and \numax. In summary, the tension between the our lowest mass measurement and other estimates is likely not due to a systematic error in one of the adopted methods, but could be due to a combined effect of small offsets in the different measurements.

Another possible explanation for the disagreement stems from the different methods we used to measure the stellar radius. Interferometry, together with the parallax--based distance, provides a measure of the Rosseland, or mean emitting surface of the star, which roughly corresponds to the point at which the optical depth $\tau = 2/3$. The radius measured from asteroseismology corresponds to the radial location where pressure waves are reflected back into the stellar interior. This occurs where the frequency of the pressure wave is smaller than the acoustic cut-off frequency, $\nu_{\rm ac}$, which for isothermal conditions depends on the sound speed and the pressure scale height (see \S~\ref{sec:background}). If the point in the stellar interior where $\nu_{\rm max} \approx \nu_{\rm ac}$ differs from the location of the $\tau = 2/3$ surface, then our two methods of measuring the stellar radius will differ.

To investigate this possibility, we examined the interior structure of a MESA model of a giant star similar to \thisstar, specifically the acoustic cutoff frequency and mean emitting surface in the outer 1\% of its radius.  We find that the $\nu_{\rm max} \approx \nu_{\rm ac}$ surface is slightly below the $\tau = 2/3$ surface. However, the difference is only 0.1\%, well below our measurement uncertainties. While this result is model--dependent, it is unlikely that the true difference is more than an order of magnitude larger than this, which is the amount required to explain the discrepancy between our various radius, and hence mass, measurements.

\section{Summary and Discussion}
\label{sec:discussion}

Our knowledge of the masses and radii of the vast majority of stars in the Galaxy rests on our theoretical understanding of stellar atmospheres and stellar evolution. For example, the process of measuring the mass of an isolated star typically begins with a measure of its effective temperature and metallicity  from its observed spectrum, and its parallax-based luminosity \citep[e.g.][]{johnson13}. For stars lacking a precise distance estimate, a spectroscopically-measured surface gravity (\logg), or its stellar density (\rhostar), can serve as a proxy for luminosity  \citep[e.g.][]{seager03,sozzetti07}. These properties define the star's location within the theoretical Herzsprung--Russell diagram, which in turn depends on the star's evolutionary state as dictated by its mass, chemical composition and age. Theoretical H--R diagrams have been computed by many groups by integrating the equations of stellar structure forward in time with various initial stellar masses and chemical compositions, and the results are tabulated in what are commonly referred to as stellar evolution model grids\footnote{These are often also referred to as ``isochrones." However, estimating stellar masses of field stars is typically performed with respect to models of fixed mass, rather than fixed age.} \citep[e.g.][]{y2}. Thus, a star's mass can be estimated by interpolating its observed properties onto these grids and recording the corresponding stellar mass, as well as other physical properties such as radius, mean density, internal structure (e.g. core helium fraction) and age \citep[e.g.][]{vf05,hekker07,takeda08, donascimento10}. 

Models of stellar atmospheres and evolution are most reliable for stars similar to the Sun, which is by far the best characterized star in the Galaxy. For locations in the H--R diagram 
that lie far from the Sun's position or for stars with different chemical compositions, theoretical atmosphere and evolution models are less robust. In these regions it is important to gather independent measurements of stellar physical characteristics that can be used to critically examine model predictions and provide touchstones for studies of stars of similar types. 


Our study focuses on an evolved star, \thisstar\, which is one of several giant stars targeted by the Doppler-based planet survey of  \citet{johnson11}. Stars such as \thisstar\ may be proxies of more massive main-sequence stars that are not amenable to precision Doppler-shift measurements owing to their rotationally-broadened absorption features. After evolving off of the main sequence, massive, hot stars shed most of their angular momentum and cool down, making them better targets for Doppler surveys. \citet{johnson07b} and \citet{johnson10c} have reported an apparent increase in the occurrence rate of giant planets around evolved stars more massive than the Sun. This has been interpreted as support for the core accretion theory of planet formation since the disks around more massive stars presumably contain more mass, and hence more of the building blocks for the protoplanetary cores that eventually become gas giants. However, the masses of these evolved stars have been called into question, raising concerns that giant stars like \thisstar, which have model-grid-based masses in excess of 1.5~\msun, may in fact have masses comparable to Sun-like dwarfs \citep{lloyd13,schlaufman13}. If this were the case, then the apparent enhanced planet occurrence rate observed around subgiant and giant stars discovered by  would require critical reexamination and perhaps a different interpretation.

There are two likely sources of systematic errors in estimating the masses of giant stars like \thisstar. The first is in the measurement of atmospheric parameters (\teff, \logg, [Fe/H]) by fitting LTE spectral models to observed spectra, which in our case is performed using the widely-used SME software package \citep{vf05}. If the effective temperature measured using this technique were off by, e.g., 200~K  then the inferred stellar mass from evolutionary models would be in error by as much as 0.3~\msun, or 20\% at the base of the red giant branch. A similar systematic error in the stellar mass would result from a 0.2~dex inaccuracy in \logg. 

Our study suggests that systematic errors in the effective temperature and surface gravity measured using SME are much smaller than 200~K and 0.2~dex, respectively. Indeed, our temperature measured from a combination of the interferometric stellar radius and bolometric luminosity agrees to well within errors with the temperature from SME. Similarly, the SME-based surface gravity agrees within errors with the asteroseismic \logg\ estimated from the observed frequency of maximum oscillation power. 

The second potential source of error in measuring the mass of giant stars is in the interpolation of the atmospheric properties onto stellar evolution model grids. Improper treatment of core overshoot, the convective mixing length parameter or other subtleties in the evolution of giant stars may lead to an inaccurate mapping of stellar physical characteristics such as mass and radius to observed properties such as luminosity, metallicity and effective temperature. We tested the veracity of the model grids  by comparing the model-grid interpolated mass and radius of \thisstar\ to the mass and radius measured from asteroseismology and interferometry, respectively. 

Interpolating the spectroscopic parameters of \thisstar\ onto the Yonsei-Yale model grids results in a mass of $\smstar \pm \usmstar$~\msun. The observed large frequency spacing and frequency of maximum oscillation observed in our \kep\ photometry yield a mass that agrees well with this estimate, giving $\amstar \pm \uamstar$~\msun. While stellar evolution models require assumptions about the complicated interplay of the interior structure of stars and the radiative transfer processes occurring in the stellar photosphere, asteroseismology provides direct measures of the bulk properties of the star, namely the mean density and surface gravity, which in turn are related to the stellar mass and radius. Thus, the agreement between our asteroseismic mass and radius and that predicted by a combination of atmospheric parameters and stellar evolution models indicates that the models are not plagued by large systematic errors.

Our observation of mixed p-- and g--modes in the oscillation spectrum of \thisstar\ provides another asteroseismic mass estimate. The period spacing of the mixed modes, $\Delta \Pi_1$, can be compared to the predictions of interior structure models (cf \S~\ref{sec:mixedmodes}). As shown in Figure~\ref{fig:dpidnu}, our observed period spacing is consistent with the mass measured from evolution model grids and asteroseismic scaling relations. These predictions, while based on interior structure models, are independent from the scaling relations used to relate the other asteroseismic parameters to stellar mass and radius. Thus, we have two independent measures of the stellar mass that agree with the mass found from model grid interpolation. 

We find some tension, at the 2.6--$\sigma$ level, to our least model-dependent mass measurement based on a combination of our interferometric radius and the density from asteroseismology. This disagreement stems primarily from a smaller radius measured from interferometry compared to the radius measured from asteroseismology. It may be that our other independent mass measurements contain independent systematic errors that result in a mass that is incorrect. Alternatively, the difference could be due to a combination of small biases in the \Dnu\ scaling relation used to derive the mean stellar density and the measured angular diameter. Additionally, there may also exist systematic errors at the $\sim 5$\% level in the model grids which contribute to this difference. It is possible that our sample of one just happens to have an interferometric mass estimate that is low due to statistical errors.

The disagreement between some of our independent mass estimates motivates further investigation using the observational techniques described herein. We are currently gathering additional asteroseismic and interferometric observations of bright, nearby evolved stars to perform a more in-depth statistical analysis of various model grids in the subgiant/giant region of the H--R diagram. However, even after adopting our smallest stellar mass estimate, we conclude that \thisstar\ has a mass that is significantly higher than that of the Sun and consistent with that an early F-- or A--type dwarf star. 

\acknowledgments
This paper includes data collected by the \kep\ mission. Funding for the Kepler mission is provided by the NASA Science Mission directorate. We are grateful to
the Kepler Team for their extensive efforts in producing such high quality data. Some of the data presented in this paper were obtained from the Multimission Archive at the Space Telescope Science Institute (MAST). STScI is operated by the Association of Universities for Research in Astronomy, Inc., under NASA contract NAS5- 26555. Support for MAST for non-HST data is provided by the NASA Office of Space Science via grant NNX09AF08G and by other grants and contracts.

Some of the data presented herein were obtained at the W.M. Keck Observatory, which is operated as a scientific partnership among the California Institute of Technology, the University of California and the National Aeronautics and Space Administration. The Observatory was made possible by the generous financial support of the W.M. Keck Foundation. We gratefully acknowledge the efforts and dedication of the Keck Observatory staff, especially Grant Hill and Scott Dahm for support of HIRES and Greg Wirth for support of remote observing. The authors wish to recognize and acknowledge the very significant cultural role and reverence that the summit of Mauna Kea has always had within the indigenous Hawaiian community. We are most fortunate to have the opportunity to conduct observations from this mountain.

JAJ is grateful for the generous grant support
provided by the Alfred P. Sloan and David \& Lucile Packard foundations, and acknowledges enlightening conversations with Dimitar Sasselov, Peter Goldreich, Phil Muirhead, Jason Wright, Debra Fischer, Jeff Valenti, James Lloyd and Victoria ``Ashley'' Villar. 

DH acknowledges support by an appointment to the NASA Postdoctoral Program at Ames Research Center administered by Oak Ridge Associated Universities, and NASA Grant NNX14AB92G issued through the Kepler Participating Scientist Program.

TSB acknowledges support provided through NASA grant ADAP12-0172.  

The CHARA Array is funded by the National Science Foundation through NSF grants AST-0606958 and AST-0908253 and by Georgia State University through the College of Arts and Sciences, as well as the W. M. Keck Foundation. 



{\it Facilities:} \facility{Shane (Hamilton Spectrograph)}, \facility{Keck (HIRES)}, \facility{Kepler}, \facility{CHARA (Classic, PAVO)}.

\clearpage

\begin{deluxetable}{lccccc}
\tabletypesize{\scriptsize}
\tablecaption{\label{tab:concordance}}
\tablewidth{0pt}
\tablehead{
\colhead{Stellar} & \colhead{LTE Spectroscopic Fit} & \colhead{Asteroseismology} & \colhead{Interferometry} &
\colhead{Asteroseismology} & \colhead{Interferometry}\\
\colhead{Parameter} & \colhead{$+$ Evolution Model\tablenotemark{a}} & \colhead{Only\tablenotemark{b}} & \colhead{and SED Fitting} & \colhead{$+$ Spectroscopy} & \colhead{$+$Asteroseismology}\\
\colhead{} & \colhead{} & \colhead{} & \colhead{} & \colhead{$+$Evolution Model\tablenotemark{c}} & \colhead{}
}
\startdata
R$_\star$ (\rsun)      &\srstar~$\pm$~\usrstar      & \arstar~$\pm$~\uarstar    & \irstar~$\pm$~\uirstar    & \asrstar~$\pm$~\uasrstar &  \ldots \\
$\rho_\star$ (\rhosun) &\srhostar~$\pm$~\usrhostar  & \arhostar~$\pm$~\uarhostar& \ldots                    & \asrhostar~$\pm$~\uasrhostar &  \ldots \\
$\log{g}$ (cgs)        &\slogg~$\pm$~\uslogg        & \alogg~$\pm$~\ualogg      & \ldots                    & \aslogg~$\pm$~\uaslogg &  \ldots \\
$T_{\rm eff}$ (K)      &\steff~$\pm$~\usteff        & \ldots                    & \iteff~$\pm$~\uiteff      & \ldots &  \ldots  \\
$[{\rm Fe/H}]$         &\sfe~$\pm$~\usfe            & \ldots                    &  \ldots                   & \ldots &  \ldots  \\
M$_\star$ (\msun)          &\smstar~~$\pm$~\usmstar     & \amstar~~$\pm$~\uamstar   & \ldots  &  \asmstar~~$\pm$~\uasmstar & \aimstar~~$\pm$~\uaimstar
\enddata

\tablenotetext{a}{Our LTE synthesis modeling was performed with SME, with \logg\ constrained using the Y$^2$ stellar evolution models. These models were also interpolated to estimate \rstar\ and \mstar.}
\tablenotetext{b}{Based on $\Dnu = \delnu \pm \sdelnu~\mu$Hz, $\numax = \nmax \pm \snmax~\mu$Hz, and Equations~\ref{eqn:dnu} and \ref{eqn:numax}.}
\end{deluxetable}

\begin{deluxetable}{lcccc}
\tabletypesize{\scriptsize}
\tablecaption{Log of interferometric observations. \label{tab:interflog}}
\tablewidth{0pt}
\tablehead{
\colhead{UT Date} & \colhead{Combiner} & \colhead{Baseline} & \colhead{No. of scans} & \colhead{Calibrators}
}
\startdata
2012 August 6 & Classic & S2-W1 & 8 & HD\,186176, HD\,188667 \\
2012 August 7 & Classic & S2-W1 & 5 & HD\,186176, HD\,188667 \\
2012 August 11 & PAVO & W1-W2 & 2 & HD\,188665 \\
2013 July 7 & PAVO & W1-W2 & 5 & HD\,177003, HD\,185872, HD\,188252 \\
2014 April 6 & PAVO & W1-W2 & 1 & HD\,185872 \\
2014 April 7 & PAVO & W1-W2 & 2 & HD\,177003, HD\,185872 \\
2014 April 10 & PAVO & E2-W2 & 1 & HD\,184784, HD\,188252
\enddata

\end{deluxetable}

\begin{deluxetable}{lcccc}
\tabletypesize{\scriptsize}
\tablecaption{Measured Angular Diameters. \label{tab:interf}}
\tablewidth{0pt}
\tablehead{
\colhead{Combiner} & \colhead{$\mu_\lambda$} & \colhead{$\theta_\mathrm{UD}$} & \colhead{$\theta_\mathrm{LD}$} & \colhead{$R$} \\
\colhead{} & \colhead{} & (mas) & (mas) & (\rsun) \\
}
\startdata
Classic & 0.32$\pm$0.04 & 1.089$\pm$0.016 & 1.120$\pm$0.018 & 4.92$\pm$0.09 \\
PAVO  & 0.64$\pm$0.03 & 1.064$\pm$0.009 & 1.133$\pm$0.013 & 4.97$\pm$0.07 \\
Classic + PAVO & \ldots & \ldots & 1.132$\pm$0.012 & 4.97$\pm$0.07
\enddata

\end{deluxetable}

\begin{deluxetable}{lcccc}
\tabletypesize{\scriptsize}
\tablecaption{Measured frequencies of \thisstar. \label{tab:frequencies}}
\tablewidth{0pt}
\tablehead{
\colhead{n\tablenotemark{a}} & \colhead{l=0} & \colhead{l=1} & \colhead{l=2} \\
\colhead{} & \colhead{($\mu$Hz)} & \colhead{($\mu$Hz)} & \colhead{($\mu$Hz)}
}
\startdata
10 &     \ldots      & 181.16$\pm$0.21 & 187.05$\pm$0.09 \\
   &                 & 182.94$\pm$0.26 & \\
11 & 188.38$\pm$0.10 & 194.24$\pm$0.09 & 202.11$\pm$0.14 \\
   &                 & 197.12$\pm$0.35 & \\
   &                 & 200.31$\pm$0.07 & \\
12 & 203.95$\pm$0.18 & 208.45$\pm$0.38 & 217.74$\pm$0.12 \\
   &                 & 211.62$\pm$0.23 & \\
13 & 219.19$\pm$0.08 & 223.10$\pm$0.12 & 232.86$\pm$0.11 \\
   &                 & 226.80$\pm$0.14 & \\
   &                 & 228.85$\pm$0.14 & \\
14 & 234.64$\pm$0.13 & 238.41$\pm$0.12 & 249.11$\pm$0.19 \\
   &                 & 242.76$\pm$0.25 & \\
   &                 & 245.06$\pm$0.05 & \\
15 & 249.83$\pm$0.17 & 257.06$\pm$0.91 & 264.56$\pm$0.17 \\
   &                 & 260.21$\pm$0.11 & \\
16 & 265.73$\pm$0.16 & 267.17$\pm$0.06 & 280.51$\pm$0.23 \\
   &                 & 272.79$\pm$0.18 & \\
   &                 & 275.57$\pm$0.65 & \\
17 & 281.46$\pm$0.29 & 288.19$\pm$0.08 & 296.06$\pm$0.36 \\
   &                 & 290.51$\pm$0.07 & \\
18 & 297.16$\pm$0.19 & 305.02$\pm$0.24 & \ldots
\enddata

\tablenotetext{a}{Value of $n$ only applies to radial ($l=0$) modes. The radial order of modes of higher degrees will be significantly different because they are mixed modes.}

\end{deluxetable}


\end{document}